\documentclass[aps,prd,numerical,showpacs,notitlepage,showkeys,10pt]{revtex4-1}

\usepackage[margin=1.8cm]{geometry}
\usepackage{amsmath}
\usepackage{amssymb}
\usepackage{amsfonts}
\usepackage{graphicx}
\usepackage{dcolumn}
\usepackage{bm}
\usepackage{hyperref}
\usepackage{wasysym}
\hypersetup{colorlinks=false}

\usepackage{graphics}
\usepackage{graphicx}
\usepackage{color}

\definecolor{LightCyan}{rgb}{0.88,1,1}
\definecolor{maroon}{cmyk}{0,0.87,0.68,0.32}
\definecolor{gray75}{gray}{0.75}
\definecolor{darkblue}{rgb}{0.089,0.21,0.363}
\definecolor{niceblue}{cmyk}{1,0.2,0.2,0.5}

\definecolor{verylightgray}{gray}{0.95}

\begin{document}

\title{Slowly-rotating curved acoustic black holes: Quasinormal modes, Hawking-Unruh radiation and quasibound states}

\date{\today}

\author{H. S. Vieira}
\email{horacio.santana.vieira@hotmail.com and horacio.santana-vieira@tat.uni-tuebingen.de}
\affiliation{Theoretical Astrophysics, Institute for Astronomy and Astrophysics, University of T\"{u}bingen, 72076 T\"{u}bingen, Germany}
\author{Kyriakos Destounis}
\email{kyriakos.destounis@uni-tuebingen.de}
\affiliation{Theoretical Astrophysics, Institute for Astronomy and Astrophysics, University of T\"{u}bingen, 72076 T\"{u}bingen, Germany}
\author{Kostas D. Kokkotas}
\email{kostas.kokkotas@uni-tuebingen.de}
\affiliation{Theoretical Astrophysics, Institute for Astronomy and Astrophysics, University of T\"{u}bingen, 72076 T\"{u}bingen, Germany}

\begin{abstract}
Astrophysical black holes are generally surrounded by accretion disks, galactic matter and the omnipresent cosmic microwave background radiation, thus allowing for the concurrent propagation of both gravitational and sound waves. Recently, acoustic black holes were embedded in Schwarzschild spacetime allowing for the coexistence of event and acoustic horizons. Here, we obtain a class of perturbative solutions to the field equations of the relativistic Gross-Pitaevskii and Yang-Mills theories, which describe sound waves propagating on a curved slowly-rotating acoustic black hole, akin to Lense-Thirring spacetime. We investigate the quasinormal mode frequencies, Hawking-Unruh radiation, and quasibound states. Our novel metric mimics the gravitational field of astrophysical compact objects in the limiting case of slow rotation, and therefore could, in principle, shed more light into the underlying classical and quantum physics of black holes through analog acoustic probes.
\end{abstract}

\maketitle

\section{Introduction}\label{Intro}

Black holes (BHs) are the most intriguing compact objects in our Universe. Not only do they play a paramount role in studying the gravitational field in extreme conditions, but also they are characterized by the most elementary macroscopic observables, namely their mass and angular momentum. The combination of simplicity together with their outstanding performance when it comes to tests of general relativity (GR) \cite{Berti:2015itd} has paved the way towards the formation of new fields of gravitational physics, such as gravitational-wave (GW) astronomy \cite{LIGOScientific:2020ibl} and BH spectroscopy \cite{Barack:2018yly}. Regardless of their observational directness, BHs undergo surprising phenomena. BHs are not completely dark when quantum effects are taken into consideration and emit Hawking radiation \cite{Hawking:1975}, with specific temperatures linked to their masses \cite{Hawking:1974}. They can also amplify incident waves under the expense of the BH's spin \cite{Penrose:1971} (or charge \cite{Bekenstein:1973,Destounis:2019hca}) if particular conditions are met, through the Penrose process \cite{Penrose:1971} and the phenomenon of superradiance \cite{Brito:2015oca}. Most importantly, astrophysical BHs are not quiet. When perturbed, they ring in accord with their characteristic properties (mass and angular momentum). The sound of BHs does not go on forever due to dissipation, which translates to a damped reverberation, known as the ringdown, associated with a set of vibrational spectra; the quasinormal mode (QNM) frequencies \cite{Kokkotas:1999bd,Berti:2009kk,Konoplya:2011qq}.

Even though the GW ringdown of BH mergers has been detected by ground-based interferometers \cite{LIGOScientific:2020ibl}, classical and quantum effects such as superradiance and Hawking radiation have not been seen in practice. To this end, different models can be used in an effort to mimic such processes that occur in the interplay between GR and controlled tabletop laboratory experiments. analog gravity \cite{Barcelo:2005fc} aims to provide insights in situations regarding phenomena that would otherwise elude observation, such as when gravitational interactions are too strong or when quantum effects become significant. The current and most promising analog gravity experiments are performed with fluids and superfluids \cite{Weinfurtner:2010nu,Richartz:2014lda,Euve:2015vml,Cardoso:2016zvz,Torres:2016iee,Patrick:2018orp,Euve:2018uyo,Patrick:2019kis,Torres:2019sbr,Torres:2020tzs,Patrick:2020baa}, Bose-Einstein condensates \cite{Garay:2000jj,Steinhauer:2014dra,MunozdeNova:2018fxv,Gooding:2020scc,Geelmuyden:2021sdh} and optical media \cite{Drori:2018ivu}, among others (see \cite{Jacquet:2020bar} and references therein). Through these remarkable experiments the phenomena of Hawking radiation \cite{Weinfurtner:2010nu,Belgiorno:2010wn}, BH superradiance \cite{Richartz:2014lda,Cardoso:2016zvz,Torres:2016iee,Patrick:2020baa}, BH ringdown and QNMs \cite{Patrick:2018orp,Torres:2019sbr,Torres:2020tzs}, as well as the cosmological redshift, Hubble friction \cite{Eckel:2017uqx} and cosmological pair creation \cite{Wittemer:2019agm} have been observed. 

A myriad of analog gravity systems share a common feature: small sound wave excitations that are described by dynamical equations of motion governing the propagation of sound in a flowing fluid can be mapped to wave equations that are encountered in classical and quantum field theory of curved
spacetimes. Such experiments stem from the pioneering work of Unruh \cite{Unruh:1980cg}, though generally a plethora of other analog experiments are employed in order to pursue a deeper understanding of strong field gravity (see \cite{Barcelo:2005fc} for a comprehensive list). In hydrodynamic experiments the sound waves experience an effective spacetime geometry that is determined by the propagation speed of these excitations and their relative speed with respect to the medium, which enables the engineering of acoustic BHs \cite{Visser:1997ux}. By tuning the velocity of the fluid, we can simulate various spacetimes; when the fluid is flowing slower than the speed of sound in one regime and faster in another, sound waves will experience the analog of a BH event horizon. The corresponding regime of the flow from which sound waves cannot escape is called a dumb hole and the surface separating causally disconnected events is called the acoustic event horizon.

The majority of the literature deals with acoustic BH geometries which are derived from $(2+1)$-dimensional Minkowski spacetime due to the lower-dimensional properties of tabletop experiments. However, acoustic BHs can be embedded in curved $(3+1)$-dimensional spacetimes as well. The motivation of such construction is threefold: (i) Astrophysical BHs reside in a bath of cosmic microwave background radiation and not in vacuum, in general. (ii) Supermassive BHs at the center of galaxies are surrounded by accretion disks and galactic halos \cite{Cardoso:2021wlq}. (iii) BHs may be further surrounded by dark matter halos consisting of quantum superfluids \cite{Berezhiani:2015bqa}. Since astrophysical BHs are usually immersed in intergalactic and cosmological media which support the propagation of sound waves, in contrast to those in vacuum, they can exhibit both an event and acoustic horizon beyond which light and sound cannot escape, respectively.

This physical perspective led to the engineering of the first acoustic Schwarzschild BH in curved spacetime \cite{Ge:2019our} by considering the field equations of the relativistic Gross-Pitaevskii \cite{Gross:1961,Pitaevskii:1961} and Yang-Mills \cite{YangMills:1954} theories. Several properties of such BHs have been analyzed, such as their QNMs, which are solutions of the linearized equations of motion with purely ingoing (outgoing) boundary conditions at the acoustic horizon (infinity), shadows \cite{Guo:2020blq} and Hawking radiation \cite{Guo:2020blq,Vieira:2021xqw}, as well as quasibound states (QBSs) \cite{Vieira:2021xqw}, which are localized mode solutions in BH potential wells that are ingoing at the acoustic event horizon and tend to zero at infinity \cite{Lasenby:2002mc,Dolan:2007mj}. More recently, this acoustic BH solution has also been extended to include charge \cite{Ling:2021vgk} in analogy with the Reissner-Nordstr\"om spacetime. 

In the present work, we utilize the approach outlined in \cite{Ge:2019our} to obtain a more astrophysically relevant effective metric that describes a slowly-rotating acoustic BH in curved spacetime, and then study the behavior of scalar fields on such spacetime. Specifically, we obtain an analog model of the Lense-Thirring BH \cite{LenseThirring} by using the Gross-Pitaevskii and Yang-Mills theories. We discuss the separability of the massless Klein-Gordon equation in this background and calculate the scalar QNMs with the Wentzel-Kramers-Brillouin (WKB) approximation \cite{Ferrari:1984zz,Schutz:1985km}. We then use the general Heun functions to analytically obtain the Hawking-Unruh radiation and the spectrum of QBSs by using the Vieira-Bezerra-Kokkotas (VBK) approach \cite{Vieira:2016ubt,Vieira:2021xqw}. In what follows, we adopt the natural units where $G = c = \hbar = 1$.

\section{The Lense-Thirring acoustic black hole}\label{LTABH}

The action of the Gross-Pitaevskii (GP) theory describing a nonlinear (complex) scalar field $\varphi$ reads
\begin{equation}
\mathcal{S}_{\rm GP}=\int d^{4}x\ \sqrt{-g}\ \biggl(|\partial_{\mu}\varphi|^{2}+m^{2}|\varphi|^{2}-\frac{b}{2}|\varphi|^{4}\biggr),
\label{eq:action_GP}
\end{equation}
where $b$ is a coupling constant and $m$ depends on the Hawking-Unruh temperature $T$ of the resultant acoustic solution which encodes both the information of BHs and the acoustic metric.
We assume the temperature dependence $m^{2} \sim T-T_{c}$, where $T_{c}$ is the critical temperature of the theory describing phase transitions with $\varphi$ being the corresponding order parameter. When $T>T_c$ the phenomenological parameter $m^2$ is positive, while for $T<T_c$ becomes negative. For $T=T_c$, $m^2=0$. 

The equation of motion with respect to $\varphi$ is given by
\begin{equation}
\Box\varphi+m^{2}\varphi-b|\varphi|^{2}\varphi=0.
\label{eq:equation_motion_GP}
\end{equation}
An acoustic BH solution of the GP theory can be obtained by considering perturbations of the complex scalar field around the spacetime background. Thus, the background spacetime metric can be fixed as
\begin{equation}
ds^{2}=g_{tt}\ dt^{2}+g_{rr}\ dr^{2}+g_{\vartheta\vartheta}\ d\vartheta^{2}+g_{\phi\phi}\ d\phi^{2}+2g_{t\phi}\ dt\ d\phi,
\label{eq2:background_LTBH_metric}
\end{equation}
where we only take into account the first order contribution of the BH's spin $a$, that is, $\mathcal{O} (a)^{2} \approx 0$, which is encoded in the cross term $g_{t\phi}$. For our purposes, this first order contribution will give rise to a term describing slowly-rotating solutions. The complex scalar field can be written in the Madelung representation as
\begin{equation}
\varphi=\sqrt{\rho(\mathbf{x},t)}\mbox{e}^{i\theta(\mathbf{x},t)},
\label{eq:Madelung_representation}
\end{equation}
where the fluid density $\rho$ and the phase $\theta$ are related to the background solution in the fixed spacetime, $(\rho_{0},\theta_{0})$, as well as to the fluctuations, $(\rho_{1},\theta_{1})$, by
\begin{eqnarray}
\rho & = & \rho_{0}+\epsilon\rho_{1},\\
\theta & = & \theta_{0}+\epsilon\theta_{1}.
\label{eq2:Madelung_representation}
\end{eqnarray}
The leading order equation for the background fluid density reads
\begin{equation}
b\rho_{0}=m^{2}-g^{\mu\nu}v_{\mu}v_{\nu},
\label{eq2:leading_order}
\end{equation}
where we have defined the background fluid four-velocity as $v_{\mu}=(-\partial_{t}\theta_{0},\partial_{i}\theta_{0})$, with $i=r,\vartheta,\phi$. At the subleading order we obtain a relativistic wave equation, similar to the massless Klein-Gordon equation, which governs the propagation of the phase fluctuations as weak excitations in a homogeneous stationary condensate. It is given by
\begin{equation}
\frac{1}{\sqrt{\mathcal{-G}}}\partial_{\mu}(\sqrt{\mathcal{-G}}\mathcal{G}^{\mu\nu}\partial_{\nu}\theta_{1})=0,
\label{eq2:phase_fluctuations}
\end{equation}
where $\mathcal{G}=\det(\mathcal{G}_{\mu\nu})$, and $\theta_{1}=\theta_{1}(t,r,\vartheta,\phi)$. The effective metric $\mathcal{G}_{\mu\nu}$, extracted from the massless Klein-Gordon equation (\ref{eq2:phase_fluctuations}), can be written as
\begin{equation}
	\mathcal{G}_{\mu\nu}=\frac{c_s}{\sqrt{c^2_s-v_{\mu}v^{\mu}}}
\begin{pmatrix}
g_{tt}(c^2_s- v_i v^i) & \vdots & -v_{t}v_{i}+g_{ti}(c_{s}^{2}-v_{i}v^{i})\cr
\cdots\cdots\cdots\cdots & \cdot &\cdots\cdots\cdots\cdots\cdots\cdots\cr
 -v_{i}v_{t}+g_{it}(c_{s}^{2}-v_{t}v^{t}) & \vdots & g_{ii}(c^2_s-v_\mu v^\mu)\delta^{ij}+v_i v_j\cr
\end{pmatrix}\!,
\label{eq2:effective_metric}
\end{equation}
where $c_{s}^{2}(\equiv b\rho_{0}/2)$ is the speed of sound. Finally, under the assumptions $v_{t} \neq 0$, $v_{r} \neq 0$, $v_{\vartheta} = 0$, $v_{\phi} = 0$, and $g_{rr}g_{tt}=-1$, with the coordinate transformations
\begin{eqnarray}
dt & \rightarrow & dt+\frac{v_{r}v_{t}}{g_{tt}(c_{s}^{2}-v_{i}v^{i})}dr,\\
d\phi & \rightarrow & d\phi-\frac{g_{t\phi}(c_{s}^{2}-v_{t}v^{t})v_{t}v_{r}}{g_{\phi\phi}(c_{s}^{2}-v_{\mu}v^{\mu})g_{tt}(c_{s}^{2}-v_{r}v^{r})}dr,
\label{eq2:coordinate_transformations}
\end{eqnarray}
we can write the line element for slowly-rotating curved acoustic BHs as
\begin{equation}
ds^{2}=c_{s}\sqrt{c_{s}^{2}-v_{\mu}v^{\mu}}\biggl(\frac{c_{s}^{2}-v_{r}v^{r}}{c_{s}^{2}-v_{\mu}v^{\mu}}g_{tt}\ dt^{2}+\frac{c_{s}^{2}}{c_{s}^{2}-v_{r}v^{r}}g_{rr}\ dr^{2}+g_{\vartheta\vartheta}\ d\vartheta^{2}+g_{\phi\phi}\ d\phi^{2}+2\frac{c_{s}^{2}-v_{t}v^{t}}{c_{s}^{2}-v_{\mu}v^{\mu}}g_{t\phi}\ dt\ d\phi\biggr).
\label{eq2:acoustic_metric}
\end{equation}
By choosing the spacetime background and the components of the fluid four-velocity, we can completely characterize the spacetime given by Eq.  \eqref{eq2:acoustic_metric}. We focus on the Lense-Thirring slowly-rotating BH background \cite{LenseThirring}, whose metric is given by \cite{MTW:1973} 
\begin{eqnarray}
ds^{2} = -f(r)\ dt^{2}+\frac{1}{f(r)}\ dr^{2}+r^{2}\ d\vartheta^{2}+r^{2}\sin^{2}\vartheta\ d\phi^{2}-\frac{4Ma\sin^{2}\vartheta}{r}\ dt\ d\phi,
\label{eq2:Lense-Thirring_metric}
\end{eqnarray}
where $a$ is the angular momentum per mass, and the metric function is $f(r)=1-2M/r$. For the four-velocity, we first rescale $m^{2} \rightarrow m^{2}/2c_{s}^{2}$ and $v_{\mu}v^{\mu} \rightarrow v_{\mu}v^{\mu}/2c_{s}^{2}$, and then by working in the limit of critical temperature of the GP theory, which implies that $m^{2}\rightarrow0$, we get $v_{\mu}v^{\mu}=-1$. Thus, the radial component of the fluid four-velocity can be chosen to be the escape velocity of an observer who maintains a stationary position at the radial position $r$. We can consider
\begin{equation}
v_{r} \sim \sqrt{\frac{2M\xi}{r}},
\label{eq2:radial_fluid_component}
\end{equation}
where $\xi$ is a tuning parameter which guarantees that the acoustic event horizon is external to the BH event horizon; this implies that the radial velocity is real for $\xi > 0$. In order to fulfill the relation $v_{\mu}v^{\mu}=-1$, the temporal component of the fluid four-velocity should be
\begin{equation}
v_{t}=\sqrt{f(r)+\frac{2M\xi}{r}f^{2}(r)}.
\label{eq2:temporal_fluid_component}
\end{equation}
Thus, we can rewrite the line element, given in Eq. (\ref{eq2:acoustic_metric}), as
\begin{equation}
ds^{2}=\sqrt{3}c_{s}^{2}\biggl[-\mathcal{F}(r)\ dt^{2}+\frac{1}{\mathcal{F}(r)}\ dr^{2}+r^{2}\ d\vartheta^{2}+r^{2}\sin^{2}\vartheta\ d\phi^{2}-\frac{4Ma\sin^{2}\vartheta}{r}\ dt\ d\phi\biggr],
\label{eq2:LTABH_metric}
\end{equation}
where the acoustic metric function $\mathcal{F}(r)$ has the following form
\begin{equation}
\mathcal{F}(r)=f(r)\biggl[1-f(r)\frac{2M\xi}{r}\biggr].
\label{eq2:f(r)_LTABH}
\end{equation}
Therefore, in the limit where $a^{2} \approx 0$ the line element describes Lense-Thirring acoustic BHs (LTABHs), with the approximation being valid up to first order in $a/r$, with $M$ the total mass at the origin. For simplicity and without loss of generality, we will fix $c_{s}^{2}=1/\sqrt{3}$. It is worth noticing that the event and acoustic horizons, at first order, do not depend on the rotation. Thus, the standard Schwarzschild BH is recovered when $\xi \rightarrow 0$. The acoustic BH covers the whole spacetime in the limit when $\xi \rightarrow \infty$, which implies that the escape velocity is such that $v_{r} \rightarrow \infty$. This is the reason for defining $\xi (> 0)$ as a tuning parameter, since it determines the position of the inner and outer acoustic event horizons.

The surface equation, given by
\begin{equation}
\mathcal{F}(r)=0=(r-r_{\rm s})(r-r_{\rm ac_{-}})(r-r_{\rm ac_{+}}),
\label{eq:surface_equation_LTABH}
\end{equation}
has three solutions, namely the BH event horizon $r_s$, as well as the inner and outer acoustic horizons $r_{\rm ac_{-}}$, $r_{\rm ac_{+}}$ of the LTABH, respectively. The BH event horizon, which is the outermost marginally trapped surface for outgoing light rays, is given by $r_{\rm s}=2M$. In turn, the inner acoustic event horizon, which corresponds to the boundary of the domain of dependence of phonon evolution, in analogy to Cauchy horizons in rotating and charged BHs, is given by $r_{\rm ac_{-}}=M(\xi-\sqrt{\xi^{2}-4\xi})$, while the outer acoustic event horizon, which is the outermost marginally trapped surface for outgoing phonons, is given by $r_{\rm ac_{+}}=M(\xi+\sqrt{\xi^{2}-4\xi})$.
An extreme LTABH is obtained in the limit $\xi \rightarrow 4$, where the inner and outer acoustic event horizons coincide at $r_{\rm ext}=4M$. 

It is worth noting that one can obtain the same effective metric given by Eq.~(\ref{eq2:LTABH_metric}), by simply extending the results obtained in \cite{Ge:2019our}, which is presented in the Appendix \ref{AppendixA}. In this alternative approach, we consider a fluid flow with circulation, which can provide some insights in the construction of an experimental apparatus to test this theory. Since the outer acoustic event horizon is the last surface from which sound waves could still escape the acoustic BH, it is meaningful to study the motion of scalar particles propagating at the external region of the LTABH spacetime, which will be presented in the next section. 

\section{Scalar wave equation}\label{SWE}

We are interested in some basic characteristics of these slowly-rotating acoustic BHs, in particular the ones related to their interaction with quantum fields, including the Hawking-Unruh radiation, and classical scalar wave scattering such as QNMs and QBSs. To this end, we have to consider the minimally coupled massless scalar field as a probe, whose covariant equation of motion is given by Eq.~(\ref{eq2:phase_fluctuations}). To obtain solutions of the massless Klein-Gordon equation, and due to stationarity and axisymmetry, we use the separation ansatz
\begin{equation}
\theta_{1}(t,r,\vartheta,\phi)=\mbox{e}^{-i \omega t}u(r)P(\vartheta)\mbox{e}^{i m \phi},
\label{eq:ansatz}
\end{equation}
where $u(r)=R(r)/r^N$ is the radial function with $N \in \mathbb{Z}$, $P(\vartheta)$ is the polar angle function, $m$ $(\in \mathbb{Z})$ is the magnetic or azimuthal quantum number, and $\omega$ is the frequency (energy, in the natural units). By substituting Eq.~(\ref{eq:ansatz}) into Eq.~(\ref{eq2:phase_fluctuations}), we obtain two ordinary differential equations
\begin{equation}
P^{\prime\prime}(\vartheta)+\frac{\cos\vartheta}{\sin\vartheta}P^{\prime}(\vartheta)+\biggl(\lambda-\frac{m^{2}}{\sin^{2}\vartheta}\biggr)P(\vartheta)=0,
\label{eq:angular_equation}
\end{equation}
\begin{equation}
R^{\prime\prime}(r)+\biggl[\frac{2(1-N)}{r}+\frac{\mathcal{F}^\prime(r)}{\mathcal{F}(r)}\biggr]R^{\prime}(r)+\frac{1}{\mathcal{F}^2(r)}\biggl\{\omega^2-\frac{4M a m \omega}{r^3}+\frac{\mathcal{F}(r)}{r^2}[N^2\mathcal{F}(r)-N\mathcal{F}(r)-Nr\mathcal{F}^\prime(r)-\lambda]\biggr\}R(r)=0,
\label{eq:radial_equation}
\end{equation}
where $\lambda$ is the separation constant and prime denotes differentiation of the polar and radial functions with respect to $\vartheta$ and $r$, respectively.

The general solution of the polar equation (\ref{eq:angular_equation}) is given in terms of the associated Legendre functions $P(\vartheta)=P_{\nu}^{m}(\cos\vartheta)$ with degree $\nu$ $(\in \mathbb{C})$ and order $m \geq 0$ $\in \mathbb{Z}$, such that $\lambda=\nu(\nu+1)$.

On the other hand, in the radial equation, which remains to be solved, the choice of $N$ is merely made for convenience. For example, $N=1$ is more convenient to study the QNMs since it enables us to write Eq. \eqref{eq:radial_equation} in a Schr\"odinger-like form, while $N=0$ is more convenient to express Eq. \eqref{eq:radial_equation} as a general Heun-type equation that is useful to obtain exact solutions, such as the Hawking-Unruh radiation and QBS spectra. 

\section{Quasinormal modes}\label{QNMs}

Even though spherical symmetry is absent in LTABHs, slow rotation enables us, in principle, to express the angular solution in terms of spherical harmonics, namely, $Y_{lm}(\vartheta,\phi) \sim P_{l}^{m}(\cos\vartheta)\mbox{e}^{i m \phi}$ with $l(=0,1,2,\ldots)$ being the angular quantum number, such that $-l \leq m \leq l$, where we used the corresponding Rodrigues' formula. Here, $\lambda \simeq l(l+1)$ approximates the separation constant in the slow rotation limit where the angular momentum operator eigenvalue equation is loosely satisfied.

By choosing $N=1$, the radial equation (\ref{eq:radial_equation}) can be written as
\begin{equation}
\frac{d^2 R(r)}{dr^2_*}+[\omega^2-V(r,\omega)]R(r)=0,
\label{QNM}
\end{equation}
with
\begin{equation}
V(r,\omega)=\frac{4M a m \omega}{r^3}+\mathcal{F}(r)\biggl[\frac{\mathcal{F}^\prime(r)}{r}+\frac{\lambda}{r^2}\biggr],
\label{eq:potential}
\end{equation}
where $dr/dr_{*}=\mathcal{F}(r)$ the tortoise coordinate. Equation \eqref{QNM} is a typical master equation \cite{Kokkotas:1999bd,Berti:2009kk} written in Schr\"odinger-like form from which we can calculate QNMs when purely ingoing (outgoing) boundary conditions are imposed at the outer acoustic horizon (infinity). Thus, by imposing purely ingoing boundary conditions at the outer acoustic horizon, $R(r)\sim e^{-i(\omega-m\Omega_1)r_*}$, and purely outgoing waves at infinity, $R(r)\sim e^{i\omega r_*}$ in Eq. \eqref{QNM}, we obtain a discrete set of QNM frequencies, characterized by the overtone number $n=0,1,2,\dots$, which differentiates between the fundamental mode $n=0$ and higher harmonics of oscillation.

There is a plethora of analytical, semianalytical and numerical schemes to calculate QNMs \cite{Kokkotas:1999bd,Berti:2009kk,Konoplya:2011qq,Hatsuda:2021gtn,Destounis:2018utr}. Even though the literature suggests the coexistence of various families of QNMs in BHs, the most common family is connected with null particles trapped at unstable circular null geodesic orbits (at the photon sphere) \cite{Cardoso:2008bp,Cardoso:2017soq,Cardoso:2018nvb,Destounis:2018qnb,Liu:2019lon,Destounis:2019omd,Destounis:2020pjk,Destounis:2020yav}. When the spacetime is perturbed the unstable null particles leak out from the photon sphere, thus giving out their preferred oscillatory states. The instability timescale of null geodesics is connected to the decay timescale of QNMs, while their oscillation frequency is connected with the angular frequency of null geodesics at the photon sphere.

In what follows, we will use the WKB approximation to calculate QNMs. The outstanding work of \cite{Ferrari:1984zz,Schutz:1985km} resulted in a derivation closely parallel to the Bohr-Sommerfeld quantization rule from quantum mechanics. The QNMs are then calculated semianalytically, using the WKB approximation \cite{Iyer:1986np,Iyer:1986nq,Kokkotas:1988fm,Seidel:1989bp,Kokkotas:1991vz,Kokkotas:1993ef}. Although it is an approximation, this approach is powerful due to its ability to be carried out in higher orders, to improve the accuracy and estimate the errors explicitly \cite{Konoplya:2003ii,Matyjasek:2017psv,Hatsuda:2019eoj,Konoplya:2019hlu}.

\begin{figure}[b]
	\centering
	\includegraphics[width=0.97\columnwidth]{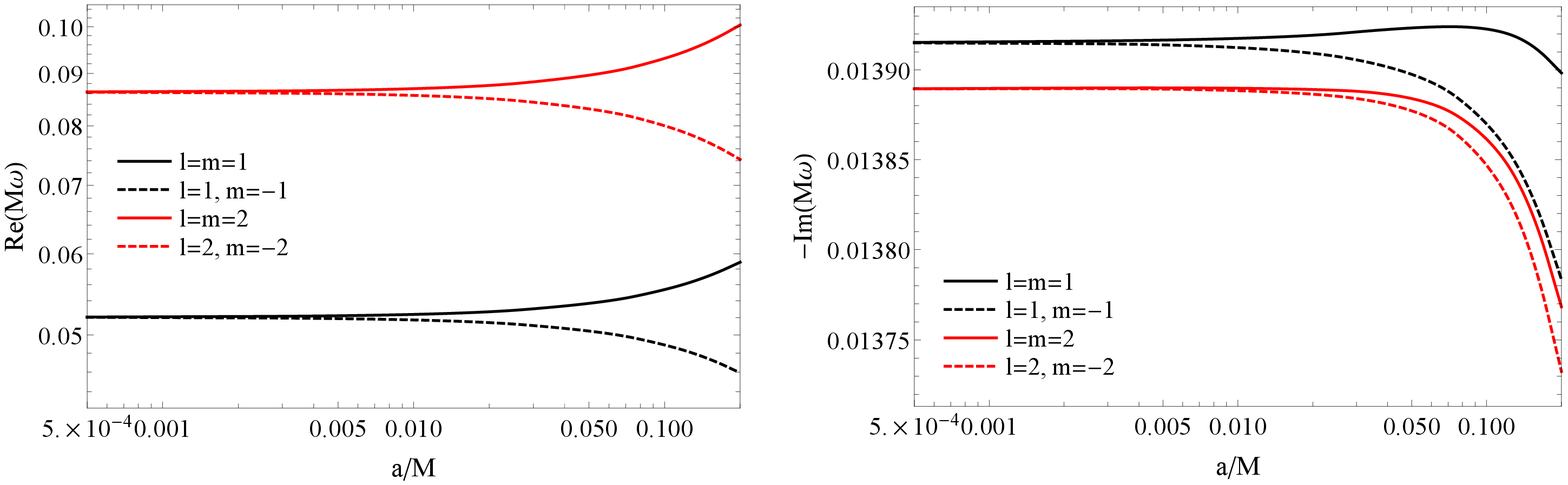}\vskip 2ex
	\includegraphics[width=0.97\columnwidth]{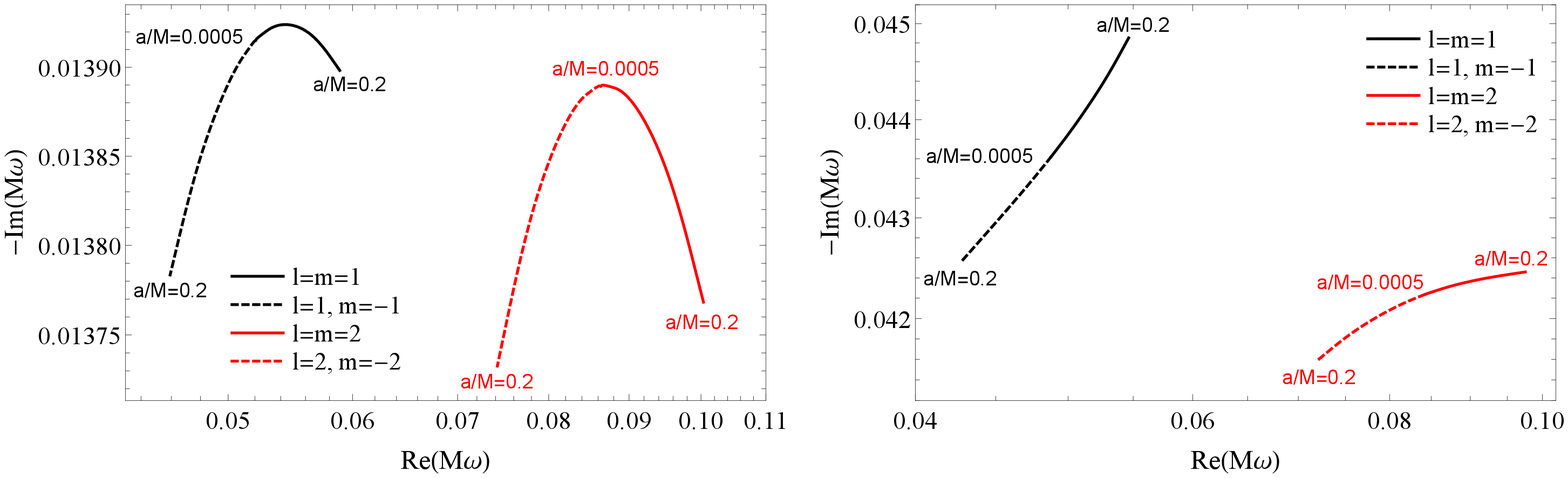}
	\caption{Top panel: real (left) and imaginary (right) parts of the fundamental $n=0$ scalar QNMs of a LTABH with $\xi=6$ and varying slow rotation parameter $a/M$. Bottom panel: fundamental $n=0$ (left) and first overtone $n=1$ (right) scalar QNM phase space of a LTABH with $\xi=6$ and varying slow rotation parameter $a/M$.}
	\label{QNMs_fixed_xi}
\end{figure}

In spherically symmetric BHs and compact objects, the effective potential of perturbations is $\omega$ independent and the WKB procedure searches for the radius $r$ for which the potential is maximized, corresponding to the photon sphere radius. After this step, the calculation of QNMs is streamlined \cite{Iyer:1986np,Konoplya:2019hlu}. Beyond spherical symmetry, or when the scalar perturbation is charged, the potential becomes $\omega$ dependent. Here, we will employ an adaption of the procedure outlined in \cite{Konoplya:2002ky} to calculate the scalar QNMs of the LTABH through Eq. \eqref{QNM}. The procedure is almost the same as for $\omega$-independent potentials though the position $r$ at which the $V(r,\omega)$ in Eq. \eqref{eq:potential} is maximized is, instead, found as a numerical function of $\omega$. The remaining procedure is identical (see \cite{Konoplya:2002ky} for further details).

Even though one cannot perform convergence tests with the WKB approximation, we have applied the same procedure to calculate QNMs for the case of Schwarzschild acoustic BHs, where $a/M\rightarrow0$, and found very good agreement with the results reported in \cite{Guo:2020blq}. We have further used the same approach to calculate the scalar QNMs of slowly-rotating Kerr BHs and our results match very well those in the known literature (e.g. \cite{Berti:2009kk,Onozawa:1996ux,Berti:2004md,Konoplya:2006br,Berti:2005ys}).

\begin{figure}[t]
	\centering
	\includegraphics[width=0.97\columnwidth]{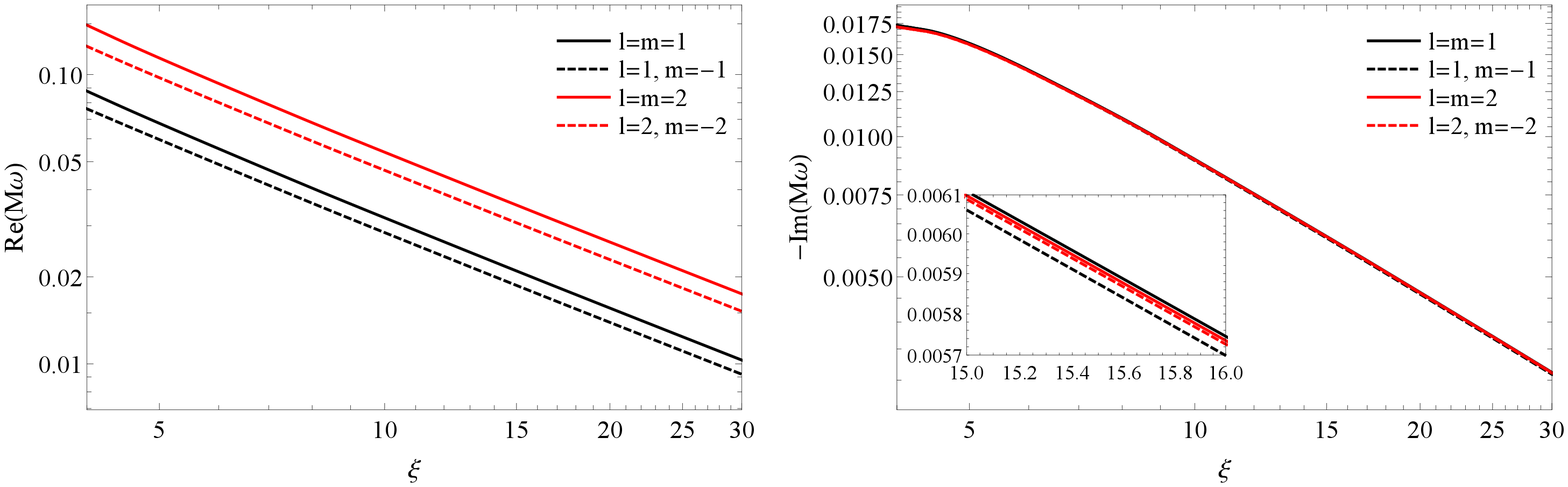}\vskip2ex
	\includegraphics[width=0.97\columnwidth]{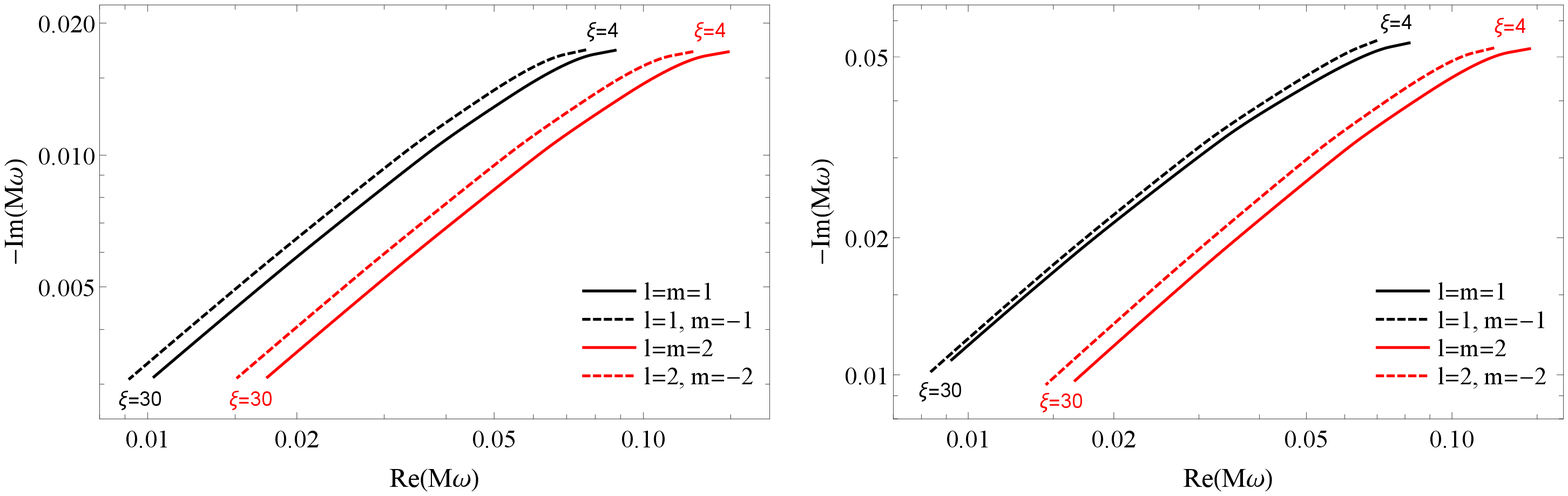}
	\caption{Top panel: real (left) and imaginary (right) parts of the fundamental $n=0$ scalar QNMs of a LTABH with $a/M=0.1$ and varying tuning parameter $\xi$. Bottom panel: fundamental $n=0$ (left) and first overtone $n=1$ (right) scalar QNM phase space of a LTABH with $a/M=0.1$ and varying tuning parameter $\xi$.}
	\label{QNMs_fixed_a}
\end{figure}

Our results are summarized in Figs. \ref{QNMs_fixed_xi} and \ref{QNMs_fixed_a}. For fixed tuning parameter $\xi$ (see Fig. \ref{QNMs_fixed_xi}), the effect of rotation on the real and imaginary parts of QNMs is similar to those found in Kerr, for which the inclusion of rotation acts like an external magnetic field on the energy levels of an atom, causing a Zeeman-like splitting of QNMs \cite{Berti:2009kk,Onozawa:1996ux,Berti:2004md,Berti:2005ys}. We also observe that the effect of rotation on the imaginary part of QNMs is rather mild, in agreement with our findings regarding QBSs (see Sec. \ref{QBSs}). On the other hand, when the rotation is fixed, the increment of $\xi$ has the tendency to decrease both the real and imaginary parts of QNMs as seen in Fig. \ref{QNMs_fixed_a}. This behavior is in complete agreement with the results in \cite{Guo:2020blq} for Schwarzschild acoustic BHs, where the angular number $l$ has a much stronger effect on the real parts rather than the imaginary ones of QNMs.

\section{Classes of exact solutions to the scalar wave equation}\label{exact_solutions}

In this section, we will analytically solve the radial part of the massless Klein-Gordon equation in the LTABH spacetime and hence present the pair of linearly independent solutions at each singularity. Furthermore, we will impose appropriate boundary conditions on the radial solution to study the Hawking-Unruh radiation and QBSs.

Thus, when $N=0$ and by using the surface equation (\ref{eq:surface_equation_LTABH}), the radial equation (\ref{eq:radial_equation}) can be written as
\begin{equation}
R^{\prime\prime}(r)+\biggl(\frac{1}{r-r_{1}}+\frac{1}{r-r_{2}}+\frac{1}{r-r_{3}}\biggr)R^{\prime}(r)+\biggl[\frac{\varpi^2-\lambda(r-r_{1})(r-r_{2})(r-r_{3})}{(r-r_{1})^{2}(r-r_{2})^{2}(r-r_{3})^{2}}\biggr]R(r)=0,
\label{eq:radial_2_LTABH}
\end{equation}
where $\varpi=r^{2}\omega-2Mam/r$. Here, we have performed a transformation on the acoustic metric function $\mathcal{F}(r) \mapsto \mathcal{F}(r)/r^{2}$, and the event horizons were renamed such that $(r_{\rm ac_{+}},r_{\rm ac_{-}},r_{\rm s})=(r_{1},r_{2},r_{3})$. Equation \eqref{eq:radial_2_LTABH} is more convenient to study QBSs with purely ingoing boundary conditions at the outer acoustic event horizon and vanishing boundary conditions at infinity, since it is a Heun-type equation \cite{Ronveaux:1995} with three finite regular singularities (associated to the three event horizons) and one regular singularity at (spatial) infinity.

Now, we follow the steps described in the VBK approach to obtain the analytical solution of Eq. \eqref{eq:radial_2_LTABH} in the LTABH spacetime (without the assumption of specific boundary conditions). First of all, we need to define a new radial coordinate, $x$, as
\begin{equation}
x=\frac{r-r_{3}}{r_{2}-r_{3}}.
\label{eq:radial_coordinate_LTABH}
\end{equation}
Then, we set a new parameter, $x_{1}$, which is associated with the three finite regular singularities by
\begin{equation}
x_{1}=\frac{r_{1}-r_{3}}{r_{2}-r_{3}}.
\label{eq:b_SABH}
\end{equation}
These definitions move the three singularities $(r_{3},r_{2},r_{1})$ to the points $(0,1,x_{1})$, while maintaining a regular singularity at (spatial) infinity. In addition, the regular singular point at $x=x_{1}$ is always located outside the unit circle $|x_{1}| > 1$, for $\xi > 4$. The last step is to perform an \textit{F-homotopic transformation} $R(x) \mapsto y(x)$ given by
\begin{equation}
R(x)=x^{A_{0}}(x-1)^{A_{1}}(x-x_{1})^{A_{x_{1}}}y(x),
\label{eq:F-homotopic}
\end{equation}
with
\begin{eqnarray}
A_{0}			& = & -\frac{i\varpi_{3}}{(r_{1}-r_{3})(r_{2}-r_{3})},\label{eq:A0}\\
A_{1}			& = & -\frac{i\varpi_{2}}{(r_{1}-r_{2})(r_{2}-r_{3})},\label{eq:A1}\\
A_{x_{1}}	& = & -\frac{i\varpi_{1}}{(r_{1}-r_{2})(r_{1}-r_{3})},\label{eq:Ax1}
\end{eqnarray}
where
\begin{equation}
\varpi_{i}=r_{i}^{2}\omega-\frac{2Mam}{r_{i}}.
\label{eq:varpi_i}
\end{equation}
Thus, by substituting Eqs.~(\ref{eq:radial_coordinate_LTABH})-(\ref{eq:varpi_i}) into Eq.~(\ref{eq:radial_2_LTABH}), we get
\begin{equation}
 y^{\prime\prime}(x)+\biggl(\frac{1+2A_{0}}{x}+\frac{1+2A_{1}}{x-1}+\frac{1+2A_{x_{1}}}{x-x_{1}}\biggr)y^{\prime}(x)+\frac{A_{3}x+A_{4}}{x(x-1)(x-x_{1})}y(x)=0,
\label{eq:radial_3_LTABH}
\end{equation}
where
\begin{eqnarray}
A_{3} & = & 2 (A_{0} A_{1}+A_{0} A_{x_{1}}+A_{0}+A_{1} A_{x_{1}}+A_{1}+A_{x_{1}})\nonumber\\
			&& +\frac{4 a m M (r_{1}^2 r_{2}+r_{1}^2 r_{3}+r_{1} r_{2}^2-6 r_{1} r_{2} r_{3}+r_{1} r_{3}^2+r_{2}^2 r_{3}+r_{2} r_{3}^2)\omega}{(r_{1}-r_{2})^2 (r_{1}-r_{3})^2 (r_{2}-r_{3})^2}\nonumber\\
			&& -\frac{2 (r_{1}^3 r_{2}^3-r_{1}^3 r_{2}^2 r_{3}-r_{1}^3 r_{2} r_{3}^2+r_{1}^3 r_{3}^3-r_{1}^2 r_{2}^3 r_{3}+3 r_{1}^2 r_{2}^2 r_{3}^2-r_{1}^2 r_{2} r_{3}^3-r_{1} r_{2}^3 r_{3}^2-r_{1} r_{2}^2 r_{3}^3+r_{2}^3 r_{3}^3)\omega^{2}}{(r_{1}-r_{2})^2 (r_{1}-r_{3})^2 (r_{2}-r_{3})^2},\label{eq:A3}\\
A_{4} & = & \frac{2 A_{0} A_{1} r_{3}-2 A_{0} A_{1} r_{1}-2 A_{0} A_{x_{1}} r_{2}+2 A_{0} A_{x_{1}} r_{3}-A_{0} r_{1}-A_{0} r_{2}+2 A_{0} r_{3}-A_{1} r_{1}+A_{1} r_{3}-A_{x_{1}} r_{2}+A_{x_{1}} r_{3}-\lambda }{r_{2}-r_{3}}\nonumber\\
			&& +\frac{4 a m M (r_{1} r_{2}+r_{1} r_{3}+r_{2} r_{3}-3 r_{3}^2)\omega}{(r_{1}-r_{3})^2 (r_{3}-r_{2})^3} +\frac{2 r_{3}^3 (r_{1} r_{3}-2 r_{1} r_{2}+r_{2} r_{3})\omega^{2}}{(r_{1}-r_{3})^2 (r_{3}-r_{2})^3}.\label{eq:A4}
\end{eqnarray}

Equation (\ref{eq:radial_3_LTABH}) has the form of a general Heun equation, whose canonical form is given by \cite{Ronveaux:1995}
\begin{equation}
y^{\prime\prime}(x)+\biggl(\frac{\gamma}{x}+\frac{\delta}{x-1}+\frac{\epsilon}{x-x_{1}}\biggr)y^{\prime}(x)+\frac{\alpha\beta x-q}{x(x-1)(x-x_{1})}y(x)=0,
\label{eq:general_Heun_canonical_form}
\end{equation}
where $y(x) \equiv \mbox{HeunG}(x_{1},q;\alpha,\beta,\gamma,\delta;x)$ denotes the general Heun function, which is the solution of Eq.~(\ref{eq:general_Heun_canonical_form}) corresponding to the exponent 0 at $x=0$ and assumes the value 1 there. If $\gamma$ is not a negative integer number, then from the Fuchs-Frobenius theory it follows that the $\mbox{HeunG}(x_{1},q;\alpha,\beta,\gamma,\delta;x)$ exists, is analytic in the disk $|x| < 1$, and has a Maclaurin expansion given by
\begin{equation}
\mbox{HeunG}(x_{1},q;\alpha,\beta,\gamma,\delta;x)=\sum_{n=0}^{\infty}c_{n}x^{n},
\label{eq:serie_HeunG_todo_x}
\end{equation}
with
\begin{eqnarray}
-qc_{0}+x_{1} \gamma c_{1} & = & 0,\nonumber\\
P_{n}c_{n-1}-(Q_{n}+q)c_{n}+X_{n}c_{n+1} & = & 0 \quad (n \geq 1),
\label{eq:recursion_General_Heun}
\end{eqnarray}
where
\begin{eqnarray}
P_{n} & = & (n-1+\alpha)(n-1+\beta),\nonumber\\
Q_{n} & = & n[(n-1+\gamma)(1+x_{1})+x_{1}\delta+\epsilon],\nonumber\\
X_{n} & = & (n+1)(n+\gamma)x_{1}.
\label{eq:P_Q_X_recursion_General_Heun}
\end{eqnarray}
Here, the normalization $c_{0}=1$ was adopted by Karl Heun. On the other hand, if $\gamma$ is not a positive integer number, the solution of Eq.~(\ref{eq:general_Heun_canonical_form}) corresponding to the exponent $1-\gamma$ at $x=0$ is $x^{1-\gamma}\mbox{HeunG}(x_{1},(x_{1}\delta+\epsilon)(1-\gamma)+q;\alpha+1-\gamma,\beta+1-\gamma,2-\gamma,\delta;x)$.

Therefore, the general, analytical solution for the radial part of the massless Klein-Gordon equation, in the LTABH spacetime, can be written as
\begin{eqnarray}
R_{j}(x)=x^{\frac{1}{2}(\gamma-1)}(x-1)^{\frac{1}{2}(\delta-1)}(x-x_{1})^{\frac{1}{2}(\epsilon-1)}[C_{1,j}\ y_{1,j}(x) + C_{2,j}\ y_{2,j}(x)],
\label{eq:analytical_solution_radial_LTABH}
\end{eqnarray}
where $C_{1,j}$ and $C_{2,j}$ are constants to be determined, and $j=\{0,1,x_{1},\infty\}$ labels the solution at each singular point. Thus, the pair of linearly independent solutions at $x=0$ ($r=r_{\rm s}$) is given by
\begin{eqnarray}
y_{1,0} & = & \mbox{HeunG}(x_{1},q;\alpha,\beta,\gamma,\delta;x),\label{eq:y10}\\
y_{2,0} & = & x^{1-\gamma}\mbox{HeunG}(x_{1},(x_{1}\delta+\epsilon)(1-\gamma)+q;\alpha+1-\gamma,\beta+1-\gamma,2-\gamma,\delta;x).\label{eq:y20}
\end{eqnarray}
Similarly, the pair of linearly independent solutions corresponding to the exponents $0$ and $1-\delta$ at $x=1$ ($r=r_{\rm ac_{-}}$) is given by
\begin{eqnarray}
y_{1,1} & = & \mbox{HeunG}(1-x_{1},\alpha\beta-q;\alpha,\beta,\delta,\gamma;1-x),\label{eq:y11}\\
y_{2,1} & = & (1-x)^{1-\delta}\mbox{HeunG}(1-x_{1},((1-x_{1})\gamma+\epsilon)(1-\delta)+\alpha\beta-q;\alpha+1-\delta,\beta+1-\delta,2-\delta,\gamma;1-x).\label{eq:y21}
\end{eqnarray}
The pair of linearly independent solutions corresponding to the exponents $0$ and $1-\epsilon$ at $x=x_{1}$ ($r=r_{\rm ac_{+}}$) is given by
\begin{eqnarray}
y_{1,x_{1}} & = & \mbox{HeunG}\biggl(\frac{x_{1}}{x_{1}-1},\frac{\alpha\beta x_{1}-q}{x_{1}-1};\alpha,\beta,\epsilon,\delta;\frac{x_{1}-x}{x_{1}-1}\biggl),\label{eq:y1x1}\\
y_{2,x_{1}} & = & \biggl(\frac{x_{1}-x}{x_{1}-1}\biggl)^{1-\epsilon}\mbox{HeunG}\biggl(\frac{x_{1}}{x_{1}-1},\frac{(x_{1}(\delta+\gamma)-\gamma)(1-\epsilon)}{x_{1}-1}+\frac{\alpha\beta x_{1}-q}{x_{1}-1};\alpha+1-\epsilon,\beta+1-\epsilon,2-\epsilon,\delta;\frac{x_{1}-x}{x_{1}-1}\biggl).\nonumber\\\label{eq:y2x1}
\end{eqnarray}
Finally, the pair of linearly independent solutions corresponding to the exponents $\alpha$ and $\beta$ at $x=\infty$ is given by
\begin{eqnarray}
y_{1,\infty} & = & x^{-\alpha}\mbox{HeunG}\biggl(\frac{1}{x_{1}},\alpha(\beta-\epsilon)+\frac{\alpha}{x_{1}}(\beta-\delta)-\frac{q}{x_{1}};\alpha,\alpha-\gamma+1,\alpha-\beta+1,\delta;\frac{1}{x}\biggl),\label{eq:y1i}\\
y_{2,\infty} & = & x^{-\beta}\mbox{HeunG}\biggl(\frac{1}{x_{1}},\beta(\alpha-\epsilon)+\frac{\beta}{x_{1}}(\alpha-\delta)-\frac{q}{x_{1}};\beta,\beta-\gamma+1,\beta-\alpha+1,\delta;\frac{1}{x}\biggl).\label{eq:y2i}
\end{eqnarray}

In these solutions, the parameters $\alpha$, $\beta$, $\gamma$, $\delta$, $\epsilon$, and $q$ are given by
\begin{eqnarray}
\alpha		& = & 1-\frac{i\varpi_{1}}{(r_{1}-r_{2})(r_{1}-r_{3})}-\frac{i\varpi_{2}}{(r_{1}-r_{2})(r_{2}-r_{3})}-\frac{i\varpi_{3}}{(r_{1}-r_{3})(r_{2}-r_{3})}+\sqrt{1-\omega^{2}},\label{eq:alpha_LTABH}\\
\beta			& = & 1-\frac{i\varpi_{1}}{(r_{1}-r_{2})(r_{1}-r_{3})}-\frac{i\varpi_{2}}{(r_{1}-r_{2})(r_{2}-r_{3})}-\frac{i\varpi_{3}}{(r_{1}-r_{3})(r_{2}-r_{3})}-\sqrt{1-\omega^{2}},\label{eq:beta_LTABH}\\
\gamma		& = & 1-\frac{2i\varpi_{3}}{(r_{1}-r_{3})(r_{2}-r_{3})},\label{eq:gamma_LTABH}\\
\delta		& = & 1-\frac{2i\varpi_{2}}{(r_{1}-r_{2})(r_{2}-r_{3})},\label{eq:delta_LTABH}\\
\epsilon	& = & 1-\frac{2i\varpi_{1}}{(r_{1}-r_{2})(r_{1}-r_{3})},\label{eq:eta_LTABH}\\
q					& = & \frac{\gamma  r_{1} \epsilon +\gamma  \delta  r_{1} x_{1}+r_{1} (-x_{1})-r_{1}-\gamma  r_{3} \epsilon -\gamma  \delta  r_{3} x_{1}+r_{3} x_{1}+r_{3}+2 \lambda  x_{1}}{2 (r_{1}-r_{3})}\nonumber\\
					&& +\frac{4 a m M x_{1} (r_{1} r_{2}+r_{1} r_{3}+r_{2} r_{3}-3 r_{3}^2)\omega}{(r_{1}-r_{3})^3 (r_{3}-r_{2})^2}
					 +\frac{2 r_{3}^3 x_{1} (-2 r_{1} r_{2}+r_{1} r_{3}+r_{2} r_{3})\omega ^2}{(r_{1}-r_{3})^3 (r_{3}-r_{2})^2}.\label{eq:q_LTABH}
\end{eqnarray}

The assumption of a specific asymptotic behavior on the aforementioned analytical solutions near the outer acoustic event horizon $r=r_1$ and spatial infinity can lead to various physical solutions. 

\subsection{Hawking-Unruh radiation}\label{Hawking-Unruh}

In this section, we analyze the Hawking-Unruh radiation spectrum of massless scalars interacting with LTABHs. To do this, we have to impose the limit $r \rightarrow r_{1}$, that is, $x \rightarrow x_{1}$, on the radial solution given by Eq.~(\ref{eq:analytical_solution_radial_LTABH}), which translates to a null argument on the general Heun functions given by Eqs.~(\ref{eq:y1x1}) and (\ref{eq:y2x1}). By following the steps described in the VBK approach, the radial solution given by Eq.~(\ref{eq:analytical_solution_radial_LTABH}), at the outer acoustic event horizon, has the following asymptotic behavior
\begin{equation}
\lim_{r \rightarrow r_{1}} R_{x_{1}}(r) \sim C_{1,x_{1}}\ (r-r_{1})^{\frac{1}{2}(\epsilon-1)} + C_{2,x_{1}}\ (r-r_{1})^{-\frac{1}{2}(\epsilon-1)},
\label{eq:asymptotic_LTABH}
\end{equation}
where $C_{1,x_{1}}$ and $C_{2,x_{1}}$ are constants to be determined, in which all remaining constants are included. Then, by including the time dependence, this solution can be written as
\begin{equation}
\Psi_{x_{1}}(r,t) \sim C_{1,x_{1}}\ \Psi_{{\rm in},x_{1}} + C_{2,x_{1}}\ \Psi_{{\rm out},x_{1}}.
\label{eq:full_wave_LTABH}
\end{equation}
where the ingoing, $\Psi_{{\rm in},x_{1}}$, and outgoing, $\Psi_{{\rm out},x_{1}}$, scalar wave solutions are given by
\begin{eqnarray}
\Psi_{{\rm in},x_{1}}(r>r_{1})	& = & \mbox{e}^{-i \omega t}(r-r_{1})^{-\frac{i(\omega-\omega_{1\rm c})}{2\kappa_{1}}},\label{eq:sol_in_LTABH}\\
\Psi_{{\rm out},x_{1}}(r>r_{1})	& = & \mbox{e}^{-i \omega t}(r-r_{1})^{+\frac{i(\omega-\omega_{1\rm c})}{2\kappa_{1}}},\label{eq:sol_out_LTABH}
\end{eqnarray}
with
\begin{equation}
\frac{1}{2}(\epsilon-1)=-\frac{i(\omega-\omega_{1\rm c})}{2\kappa_{1}}.
\label{eq:gamma_wave_LTABH}
\end{equation}
Here, $\omega_{1\rm c}=m\Omega_{1}$ is the critical value of the frequency at the outer acoustic event horizon, which can be associated with the effect of superradiance. The effect will occur when $0 < \omega_{R} < \omega_{1\rm c}$, where $\omega_{R}$ is the real part of the frequency $\omega$. However, in our case, since we are dealing with a slowly-rotating acoustic BH, with $a \ll 1$, this critical value will not be reached, though in principle, for large values of $m$ it should be possible even if $a \ll 1$. In addition, the gravitational acceleration, $\kappa_{1}$, and the angular velocity of the dragging of inertial frames, $\Omega_{1}$, on the outer acoustic event horizon are given by
\begin{eqnarray}
\kappa_{1} & \equiv & \frac{1}{2r_{1}^{2}} \left.\frac{d\mathcal{F}(r)}{dr}\right|_{r=r_{1}} = \frac{(r_{1}-r_{2})(r_{1}-r_{3})}{2r_{1}^{2}},\label{eq:grav_acc_LTABH}\\
\Omega_{1} & \equiv & -\left.\frac{g_{t\phi}}{g_{\phi\phi}}\right|_{r=r_{1}} = \frac{2Ma}{r_{1}^{3}}.\label{eq:dragging_LTABH}
\end{eqnarray}

Therefore, by using the analytic continuation described in \cite{Vieira:2014waa}, we can obtain the relative scattering probability, $\Gamma_{1}$, and the exact spectrum of Hawking-Unruh radiation, $\bar{N}_{\omega}$, given by
\begin{eqnarray}
\Gamma_{1}				& = & \left|\frac{\Psi_{{\rm out},x_{1}}(r>r_{1})}{\Psi_{{\rm out},x_{1}}(r<r_{1})}\right|^{2}=\mbox{e}^{-\frac{2\pi(\omega-\omega_{1\rm c})}{\kappa_{1}}},\label{eq:rel_prob_LTABH}\\
\bar{N}_{\omega}	& = & \frac{\Gamma_{1}}{1-\Gamma_{1}}=\frac{1}{\mbox{e}^{(\omega-\omega_{1\rm c})/k_{B}T_{1}}-1}.\label{eq:rad_spec_LTABH}
\end{eqnarray}
Here, $k_{B}$ is the Boltzmann constant, and $T_{1}(=\kappa_{1}/2\pi k_{B})$ is the Hawking-Unruh temperature at the outer acoustic event horizon. The resulting spectrum of Hawking-Unruh radiation, for massless scalars in the LTABH spacetime, has a thermal character and hence it is analogous to the spectrum of black body radiation. 

\subsection{Quasibound states}\label{QBSs}

The QBSs (also known as resonance spectra \cite{Damour:1976kh} or quasistationary levels \cite{Gaina:1993ib}) are solutions of the equation of motion \eqref{eq:radial_2_LTABH}, localized at the BH potential well which tends to zero at spatial infinity. The spectrum of QBSs is inevitably complex due to dissipation of energy at the outer acoustic event horizon thus can be expressed as $\omega=\omega_{R}+i\omega_{I}$, where $\omega_{R}$ is the real (oscillation frequency) part, and $\omega_{I}$ is the imaginary (decay or growth rate) part.
This means that one has to impose two boundary conditions on the radial solution, namely, ingoing waves at the outer acoustic event horizon, while at spatial infinity they should vanish. In order to obtain the spectrum of QBSs, we follow the VBK approach, which suggests imposing the polynomial condition of the general Heun functions as a matching condition for the two asymptotic behaviors of the radial solution. From here on in this section, we set $M=1$.

From the asymptotic behavior at the outer acoustic event horizon described by Eq.~(\ref{eq:full_wave_LTABH}), we must impose that $C_{2,x_{1}}=0$ in order to satisfy purely ingoing boundary conditions. Let us now analyze under which circumstances the radial solution vanishes at spatial infinity. To do this, we use the two linearly independent solutions of the general Heun equation at spatial infinity, given by Eqs.~(\ref{eq:y1i}) and (\ref{eq:y2i}), to obtain the following asymptotic behavior for the radial solution given by Eq.~(\ref{eq:analytical_solution_radial_LTABH}),
\begin{equation}
\lim_{r \rightarrow \infty} R_{\infty}(r) \sim C_{1,\infty}\ \frac{1}{r^{\sigma}},
\label{eq:radial_infinity_LTABH}
\end{equation}
where $\sigma = - i A + i B \omega + \alpha$. The coefficients $A$, and $B$ are real numbers, given by
\begin{eqnarray}
A & = & \frac{\omega_{1\rm c}}{2\kappa_{1}}-\frac{\omega_{2\rm c}}{2\kappa_{2}}+\frac{\omega_{3\rm c}}{2\kappa_{3}} = \frac{am[\xi(\xi-4)+(\xi-3)\sqrt{\xi(\xi-4)}]}{4\xi(\xi-4)},\label{eq:A_LTABH}\\
B & = & \frac{1}{2\kappa_{1}}-\frac{1}{2\kappa_{2}}+\frac{1}{2\kappa_{3}} = 1+\frac{2\xi}{\sqrt{\xi(\xi-4)}},\label{eq:B_LTABH}
\end{eqnarray}
where $\omega_{2\rm c}=m\Omega_{2}$ and $\omega_{3\rm c}=m\Omega_{3}$ are the critical values of the frequency $\omega$ at the inner acoustic and BH event horizons, respectively. Here, the gravitational accelerations, $\kappa_{2}$ and $\kappa_{3}$, and the angular velocities, $\Omega_{2}$ and $\Omega_{3}$, can be obtained in a similar manner as in Eqs.~(\ref{eq:grav_acc_LTABH}) and (\ref{eq:dragging_LTABH}), respectively. In the nonrotating case, that is, when $a=0$, we get $A=0$ and recover the parameter $\sigma$ related to Schwarzschild acoustic BHs \cite{Vieira:2021xqw}. In addition, since we are focusing on the subextremal case ($\xi > 4$), it is worth noticing that the coefficient $B$ is positive definite, such that $B > 1$. Furthermore, the sign of coefficient $A$ depends on the sign of $m$, which shows the direction of rotation. Therefore, the radial solution fully satisfies the second boundary condition for QBSs if the sign of the real part of $\sigma$ is such that $\mbox{Re}[\sigma] > 0$; whereas the radial solution diverges at spatial infinity if $\mbox{Re}[\sigma] < 0$.

The asymptotic behavior of the radial solution at spatial infinity will be determined when we know the values of the coefficients $A$ and $B$, as well as the frequencies $\omega$, and the parameter $\alpha$. This will be obtained as follows. We will use a polynomial condition for the general Heun functions to match the two asymptotic solutions of the scalar radial equation in a common overlap region. A general Heun function becomes a (class I) polynomial of degree $n$ if it satisfies the following (necessary, but not sufficient) condition \cite{Ronveaux:1995}:
\begin{equation}
\alpha=-n,
\label{eq:alpha-condition}
\end{equation}
where $n=0,1,2,\ldots$. In fact, it is worth mentioning that only the general Heun functions of class I are polynomials in the strict sense, which are solutions of the general Heun equation valid at the three finite regular singularities, in the sense of being simultaneously a Frobenius solution at each of these singularities. Furthermore, it can be shown that if a general Heun polynomial exists, which is true in our case, then it must also be a Frobenius solution about the regular singularity at spatial infinity, and hence it is analytic in the whole finite $x$ plane, except at the singularities and with appropriate cuts made to ensure single valuedness. In addition to this so-called $\alpha$ condition, the accessory parameter $q$ must be appropriately chosen so that Eqs. (\ref{eq:recursion_General_Heun}) and (\ref{eq:P_Q_X_recursion_General_Heun}) are consistent, which means that it is also necessary for the accessory parameter $q$ to be an eigenvalue of the general Heun equation, calculated via the polynomial equation $c_{n+1}=0$. Indeed, the accessory parameter $q$ contains the separation constant $\lambda$, which indicates that we could obtain the eigenvalues of the separation constant $\lambda$, corresponding to the appropriate eigenvalues of the frequencies $\omega$, from the polynomial solution for the radial equation and then use it to show the (regular) angular behavior of the massless scalar QBSs in the LTABH spacetime.

In the present case, the parameter $\alpha$, which is given by Eq.~(\ref{eq:alpha_LTABH}), can be written as
\begin{equation}
\alpha=1+iA-iB\omega+\sqrt{1-\omega^{2}}.
\label{eq:alpha_2_LTABH}
\end{equation}
Thus, by imposing the $\alpha$ condition given by Eq.~(\ref{eq:alpha-condition}), we obtain a second order equation for $\omega$, whose solutions form the exact spectrum of QBSs:
\begin{equation}
\omega_{mn}^{(\pm)} = \frac{-i(n+1)B + AB \pm \sqrt{A^{2}-2i(n+1)A-n(n+2)-B^{2}}}{B^{2}-1},
\label{eq:omega_LTABH}
\end{equation}
where $n$ is the overtone number, which, without loss of generality, can be called also the principal quantum number. Note that in the nonrotating case, when $a=0$, which implies that $A=0$ (the same can be said when $m=0$), we recover the solutions concerning the Schwarzschild acoustic BH spacetime \cite{Vieira:2021xqw}; in this case, the minus solution is the physically admissible one.

In Tables \ref{tab:I_LTABH} and \ref{tab:II_LTABH} we present the QBSs $\omega_{mn}^{(-)}$ and $\omega_{mn}^{(+)}$, respectively, as functions of the tuning parameter $\xi$, for slow rotation $a/M$. We can conclude that the minus solution is physically admissible for $m=0,-1,-2,\ldots$, as well as the plus solution for $m=+1,+2,\ldots$, since the real part of $\sigma$ is positive in these cases, and therefore they represent QBS frequencies for massless scalars in the LTABH spacetime. Furthermore, by comparing our Tables \ref{tab:I_LTABH} and \ref{tab:II_LTABH} with Table I in \cite{Vieira:2021xqw}, we observe that the imaginary parts of both minus and plus solutions in the LTABH spacetime are equal to the those in the Schwarzschild acoustic BH spacetime, thus the introduction of a menial rotation ($\sim 10^{-3}$) does not significantly affect the decay rate of QBSs. Motivated by the slow rotation approximation of Kerr perturbations and the fact that they match well the QNMs of the full subextremal problem, it is meaningful to analyze QBSs for slightly larger values of angular momentum $a$, up to $20\%$ -- $30\%$ of the Kerr rotation.

\begin{table}[t]
	\caption{The massless scalar QBSs $\omega_{mn}^{(-)}$, and the corresponding values of the real part of $\sigma_{mn}^{(-)}=-iA+i B \omega_{mn}^{(-)}-n$, for $a/M=0.001$. We focus on the modes with $n=0,\,1$ and $m=\pm 1$.}
	\label{tab:I_LTABH}
	\begin{tabular}{c||c|c||c|c||c|c||c|c}
		\hline\noalign{\smallskip}
		$\xi$    & $\omega_{-10}^{(-)}$  & $\sigma_{-10}^{(-)}$ & $\omega_{+10}^{(-)}$ & $\sigma_{+10}^{(-)}$ & $\omega_{-11}^{(-)}$       & $\sigma_{-11}^{(-)}$ & $\omega_{+11}^{(-)}$ & $\sigma_{+11}^{(-)}$ \\
		\noalign{\smallskip}\hline\noalign{\smallskip}
		4.01     & $-0.000036-0.048750i$ & $2.001190$           & $0.000036$           & $0$                  & $-0.000036-0.073146i$      & $2.002670$           & $0.000036-0.024353i$ & $-0.000296$ \\
		5        & $-0.000093-0.378115i$ & $2.069098$           & $0.000087$           & $0$                  & $-0.000095-0.576417i$      & $2.154234$           & $0.000084-0.179813i$ & $-0.016038$ \\
		6        & $-0.000116-0.471688i$ & $2.105662$           & $0.000105$           & $0$                  & $-0.000120-0.724662i$      & $2.234963$           & $0.000100-0.218714i$ & $-0.023639$ \\
		7        & $-0.000130-0.525149i$ & $2.129505$           & $0.000115$           & $0$                  & $-0.000137-0.810673i$      & $2.287319$           & $0.000109-0.239625i$ & $-0.028309$ \\
		8        & $-0.000141-0.560660i$ & $2.146447$           & $0.000123$           & $0$                  & $-0.000148-0.868345i$      & $2.324395$           & $0.000116-0.252975i$ & $-0.031502$ \\
		9        & $-0.000149-0.586203i$ & $2.159152$           & $0.000129$           & $0$                  & $-0.000157-0.910095i$      & $2.352136$           & $0.000120-0.262312i$ & $-0.033831$ \\
		10       & $-0.000155-0.605544i$ & $2.169052$           & $0.000133$           & $0$                  & $-0.000164-0.941855i$      & $2.373714$           & $0.000124-0.269233i$ & $-0.035609$ \\
		$\vdots$ & $\vdots$              & $\vdots$             & $\vdots$             & $\vdots$             & $\vdots$                   & $\vdots$             & $\vdots$               & $\vdots$    \\
		$\infty$ & $-0.000208-0.750000i$ & $2.250000$           & $0.000166$           & $0$                  & $-0.000223-1.183010i$      & $2.549040$           & $0.000151-0.316987i$ & $-0.049038$ \\
		\noalign{\smallskip}\hline
	\end{tabular}
\end{table}

\begin{table}[t]
	\caption{The massless scalar QBSs $\omega_{mn}^{(+)}$, and the corresponding values of the real part of $\sigma_{mn}^{(+)}=-iA+i B \omega_{mn}^{(+)}-n$, for $a/M=0.001$. We focus on the modes with $n=0,\,1$ and $m=\pm 1$.}
	\label{tab:II_LTABH}
	\begin{tabular}{c||c|c||c|c||c|c||c|c}
		\hline\noalign{\smallskip}
		$\xi$    & $\omega_{-10}^{(+)}$ & $\sigma_{-10}^{(+)}$ & $\omega_{+10}^{(+)}$ & $\sigma_{+10}^{(+)}$ & $\omega_{-11}^{(+)}$        & $\sigma_{-11}^{(+)}$ & $\omega_{+11}^{(+)}$ & $\sigma_{+11}^{(+)}$ \\
		\noalign{\smallskip}\hline\noalign{\smallskip}
		4.01     & $-0.000036$          & $0$                  & $0.000036-0.048750i$ & $2.001190$           & $-0.000036-0.024353i$       & $-0.000296$          & $0.000036-0.073146i$ & $2.002670$ \\
		5        & $-0.000087$          & $0$                  & $0.000093-0.378115i$ & $2.069098$           & $-0.000084-0.179813i$       & $-0.016038$          & $0.000095-0.576417i$ & $2.154234$ \\
		6        & $-0.000105$          & $0$                  & $0.000116-0.471688i$ & $2.105662$           & $-0.000100-0.218714i$       & $-0.023639$          & $0.000120-0.724662i$ & $2.234963$ \\
		7        & $-0.000115$          & $0$                  & $0.000130-0.525149i$ & $2.129505$           & $-0.000109-0.239625i$       & $-0.028309$          & $0.000137-0.810673i$ & $2.287319$ \\
		8        & $-0.000123$          & $0$                  & $0.000141-0.560660i$ & $2.146447$           & $-0.000116-0.252975i$       & $-0.031502$          & $0.000148-0.868345i$ & $2.324395$ \\
		9        & $-0.000129$          & $0$                  & $0.000149-0.586203i$ & $2.159152$           & $-0.000120-0.262312i$       & $-0.033831$          & $0.000157-0.910095i$ & $2.352136$ \\
		10       & $-0.000133$          & $0$                  & $0.000155-0.605544i$ & $2.169052$           & $-0.000124-0.269233i$       & $-0.035609$          & $0.000164-0.941855i$ & $2.373714$ \\
		$\vdots$ & $\vdots$             & $\vdots$             & $\vdots$             & $\vdots$             & $\vdots$                    & $\vdots$             & $\vdots$             & $\vdots$   \\
		$\infty$ & $-0.000166$          & $0$                  & $0.000208-0.750000i$ & $2.250000$           & $-0.000151-0.316987i$       & $-0.049038$          & $0.000223-1.183010i$ & $2.549040$ \\
		\noalign{\smallskip}\hline
	\end{tabular}
\end{table}

\begin{figure}[t]
	\centering
	\includegraphics[width=0.5\columnwidth]{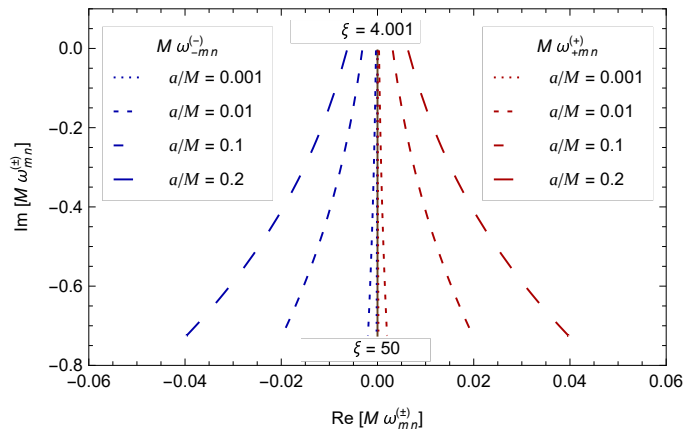}
	\caption{The massless scalar QBSs $\omega_{mn}^{(\pm)}$ for $n=0$, $|m|=1$, and different values of the angular momentum $a/M$, as the tuning parameter $\xi$ runs from $4.001$ (on the top) to $50$ (on the bottom). The brown straight line represents the nonrotating case $a/M=0$, where QBSs are purely imaginary.}
	\label{fig:QBSs}
\end{figure}

In Table \ref{tab:III_LTABH} we see that the slow rotation affects the field energies by decreasing (increasing) the real (imaginary) part of the QBS frequencies in the LTABH spacetime, for a fixed value of the tuning parameter $\xi$. As it was expected, the imaginary part $\mbox{Im}[\omega_{mn}^{(\pm)}]$ is only slightly shifted. In addition, we also present the behavior of the QBSs $\omega_{mn}^{(\pm)}$ in Fig.~\ref{fig:QBSs}, as functions of the tuning parameter $\xi$, for some values of the angular momentum $a/M$.

\begin{table}[t]
	\caption{The physically admissible massless scalar QBSs $\omega_{-mn}^{(-)}$ and $\omega_{+mn}^{(+)}$, for $|m|=1$, $\xi=5$, and different values of the angular momentum $a/M$.}
	\label{tab:III_LTABH}
	\begin{tabular}{c|c|c}
		\hline\noalign{\smallskip}
		$a/M$  & $\omega_{-10}^{(-)}$  & $\omega_{+10}^{(+)}$ \\
		\noalign{\smallskip}\hline\noalign{\smallskip}
		0.01 & $-0.000925-0.378115i$ & $0.000925-0.378115i$ \\
		0.05 & $-0.004626-0.378114i$ & $0.004626-0.378114i$ \\
		0.10 & $-0.009253-0.378108i$ & $0.009253-0.378108i$ \\
		0.15 & $-0.013879-0.378100i$ & $0.013879-0.378100i$ \\
		0.20 & $-0.018506-0.378088i$ & $0.018506-0.378088i$ \\
		0.25 & $-0.023132-0.378073i$ & $0.023132-0.378073i$ \\
		0.30 & $-0.027759-0.378054i$ & $0.027759-0.378054i$ \\
		\noalign{\smallskip}\hline
	\end{tabular}
\end{table}

We conclude that the massless scalar QBS frequencies $\omega_{mn}^{(\pm)}$ in a LTABH spacetime satisfy the following symmetry,
\begin{equation}
\omega_{mn}^{(+)}=-[\omega_{-mn}^{(-)}]^{*},
\label{eq:resonant_frequencies_symmetry}
\end{equation}
where ``$^{*}$'' denotes complex conjugation. This symmetry indicates that these two solutions have the same decay rate, $\mbox{Im}[\omega_{mn}^{(+)}]=\mbox{Im}[\omega_{-mn}^{(-)}]$, and opposite oscillation frequency signs, $\mbox{Re}[\omega_{mn}^{(+)}]=-\mbox{Re}[\omega_{-mn}^{(-)}]$, which implies a symmetry under the transformation $m \rightarrow -m$. This symmetry may describe the simultaneous particle-antiparticle creation under phonon interaction \cite{Deruelle:1974PL}.

\subsection{Radial wave eigenfunctions of quasibound states}\label{Radial_eigenfunctions}

In this section, we present the radial wave eigenfunctions related to QBSs of massless scalars propagating in the LTABH background. To do this, we also use the VBK approach (for details, see Refs.~\cite{Vieira:2016ubt,Vieira:2021xqw}). As we explained before, these polynomial eigenfunctions are related to the appropriate determination of the accessory parameter $q$; since it is calculated via the polynomial equation $c_{n+1}=0$, we index its solutions by a parameter $s$ running from $0$ to $n$, which can be conveniently denoted by $q_{mn;s}$.

Therefore, the QBS radial wave eigenfunctions for massless scalars propagating in a LTABH spacetime are given by
\begin{equation}
R_{mn;s}^{(\pm)}(x)=C_{mn;s}^{(\pm)}\ x^{\frac{1}{2}(\gamma-1)}(x-1)^{\frac{1}{2}(\delta-1)}(x-x_{1})^{\frac{1}{2}(\epsilon-1)}\ \mbox{HeunGp}_{mn;s}^{(\pm)}(x_{1},q_{mn;s};-n,\beta,\gamma,\delta;x),
\label{eq:eigenfunctions_LTABH}
\end{equation}
where $C_{mn;s}^{(\pm)}$ is a constant to be determined, $\mbox{HeunGp}_{mn;s}^{(\pm)}(x_{1},q_{mn;s};-n,\beta,\gamma,\delta;x)$ are the general Heun polynomials, the signs $(\pm)$ are related to $\omega_{mn}^{(\pm)}$, and $s$ labels the degeneracy of the eigenfunctions.

\begin{figure}[t]
	\centering
	\includegraphics[width=1\columnwidth]{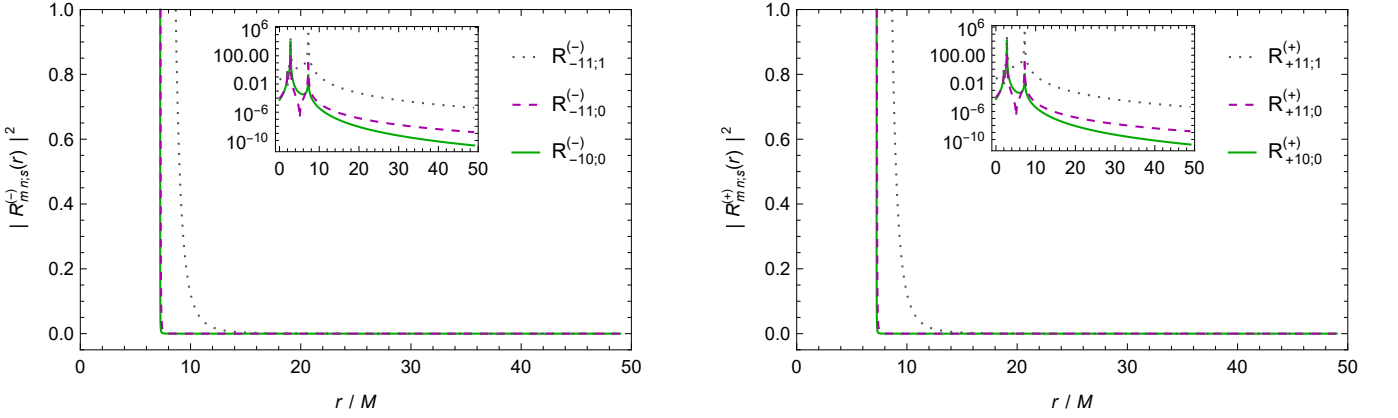}
	\caption{Left: the first three squared QBS radial wave eigenfunctions related to $\omega_{mn}^{(-)}$, for $n=0$, $m=-1$, $\xi=5$, and $a/M=0.1$. The units are in multiples of $C_{mn;s}^{(-)}$. Right: the first three squared QBS radial wave eigenfunctions related to $\omega_{mn}^{(+)}$, for $n=0$, $m=+1$, $\xi=5$, and $a/M=0.1$. The units are in multiples of $C_{mn;s}^{(+)}$.}
	\label{fig:Eigenfunctions}
\end{figure}

In Fig.~\ref{fig:Eigenfunctions}, we plot the first three squared radial wave eigenfunctions. We observe that the radial solution tends to zero at spatial infinity and diverges at the outer acoustic event horizon, which represents QBSs. Note that the radial eigenfunctions reach a maximum value (at the outer acoustic horizon $r_{1}=7.236070$, for $\xi=5$ and $a/M=0.1$) and then cross into the BH, as shown in the inlay log-scale plots.

\subsection{Angular wave eigenfunctions of quasibound states}\label{Angular_eigenfunctions}

It is known that the regularity requirements of the angular functions at the two boundaries $\vartheta=0$ and $\vartheta=\pi$ single out a discrete set of angular eigenvalues $\lambda$, which couples the angular equation (\ref{eq:angular_equation}) and the radial equation (\ref{eq:radial_equation}). In the present case, we can obtain an expression for the eigenvalues $\lambda$ from the polynomial equation $c_{n+1}=0$ that determines the accessory parameter $q_{mn;s}$.

Therefore, for the fundamental QBS $n=0$, the eigenvalues $q_{m0;s}$ must obey the relation $c_{1}=0$, where $c_{1}=q/x_{1}\gamma$, and then we have that $q_{m0;0}=0$; in this case $s=0$ because we get only one solution for $q$. Thus, from Eq.~(\ref{eq:q_LTABH}), we obtain the following expression for the eigenvalues
\begin{eqnarray}
\lambda_{m0}^{(\pm)} & = & -\frac{i a m \{5 \sqrt{(\xi -4) \xi }+\xi \{\xi [\xi +\sqrt{(\xi -4) \xi }-6]-5 \sqrt{(\xi -4) \xi }+8\}\}}{2 (\xi -4) \xi }\nonumber\\
				&& +\frac{a m \{3 \sqrt{(\xi -4) \xi }+\xi [7-2 \xi -2 \sqrt{(\xi -4) \xi }]-5\}+2 i (\xi -2) [\xi +\sqrt{(\xi -4) \xi }]}{\sqrt{(\xi -4) \xi }}\omega_{m0}^{(\pm)}\nonumber\\
				&& +4 \biggl[\frac{(\xi -2) \xi }{\sqrt{(\xi -4) \xi }}+\xi \biggr][\omega_{m0}^{(\pm)}]^{2}.
\label{eq:lambda}
\end{eqnarray}
As it was expected, the eigenvalues $\lambda_{m0}^{(\pm)}$ are complex; the application of complex angular momentum techniques to atomic and molecular scattering by evaluating the Legendre functions of the first kind of complex degree $\nu$ was reported in the end of the 1970s \cite{Connor:1979MP}. Thus, the complex degree $\nu$ can be numerically evaluated. 

\begin{table}[b]
	\caption{The eigenvalue $\lambda_{m0}^{(\pm)}$, and the corresponding complex degree $\nu_{m0}^{(\pm)}$, related to the physically admissible massless scalar QBSs $\omega_{-m0}^{(-)}$ and $\omega_{+m0}^{(+)}$, for $\xi=5$, and $a/M=0.1$.}
	\label{tab:IV_LTABH}
	\begin{tabular}{c||c|c||c|c}
		\hline\noalign{\smallskip}
		$m$	&	$\lambda_{-m0}^{(-)}$	&	$\nu_{-m0}^{(-)}$			&	$\lambda_{+m0}^{(+)}$	&	$\nu_{+m0}^{(+)}$			\\
		\noalign{\smallskip}\hline\noalign{\smallskip}
		0		&	$0.645898$						&	$0.446519$						&	$0$										&	0											\\
		1		&	$0.635264-0.193022i$	&	$0.446395-0.101978i$	&	$0.635264+0.193022i$	&	$0.446395+0.101978i$	\\
		2		&	$0.603362-0.386013i$	&	$0.446036-0.204016i$	&	$0.603362+0.386013i$	&	$0.446036+0.204016i$	\\
		3		&	$0.550192-0.578943i$	&	$0.445478-0.306164i$	&	$0.550192+0.578943i$	&	$0.445478+0.306164i$	\\
		4		&	$0.475753-0.771779i$	&	$0.444767-0.408450i$	&	$0.475753+0.771779i$	&	$0.444767+0.408450i$	\\
		5		&	$0.380045-0.964492i$	&	$0.443951-0.510880i$	&	$0.380045+0.964492i$	&	$0.443951+0.510880i$	\\
		\noalign{\smallskip}\hline
	\end{tabular}
\end{table}

\begin{figure}[t]
	\centering
	\includegraphics[width=0.5\columnwidth]{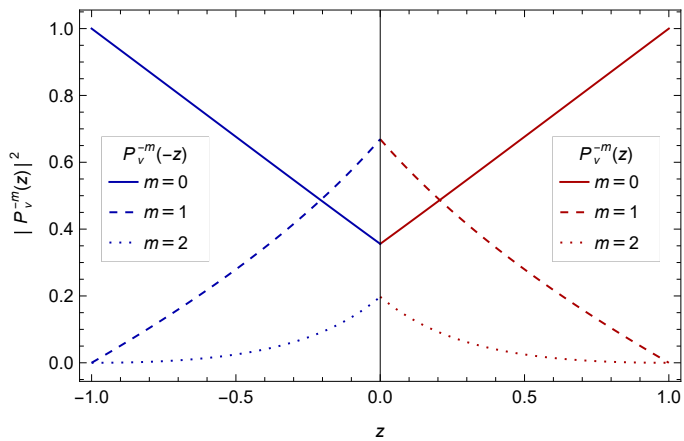}
	\caption{The first three squared QBS angular eigenfunctions related to $\omega_{-m0}^{(-)}$, for $\xi=5$, and $a/M=0.1$, where $z=\cos\vartheta$.}
	\label{fig:Angular}
\end{figure}

In Table \ref{tab:IV_LTABH}, we present the eigenvalues $\lambda_{m0}^{(\pm)}$, as well as the corresponding complex degree $\nu_{m0}^{(\pm)}$, as a function of the azimuthal quantum number $m$, for fixed values of the tuning parameter $\xi$ and angular momentum $a/M$. In addition, we also present the behavior of the QBS angular eigenfunctions $P(\vartheta)$ in Fig.~\ref{fig:Angular}, as functions of the new polar coordinate $z=\cos\vartheta$, for some values of the angular momentum $a/M$. It is worth emphasizing that the numerically satisfactory solutions of the associated Legendre equation of complex degree are given in terms of the Ferrers function of the first kind $P_{\nu}^{-m}(-z)$ and $P_{\nu}^{-m}(z)$ in the interval $-1 < x < 1$. Note that these angular solutions are regular at the two boundaries $\vartheta=\pi$ ($z=-1$) and $\vartheta=0$ ($z=1$).

\section{Conclusions}\label{VII}

In this work, we obtained a new effective metric describing a slowly-rotating and curved acoustic BH spacetime. This solution is an analog model of the well-known Lense-Thirring BH, which can be supported in astrophysical scenarios under the assumption that the BHs in our Universe are immersed in galactic and cosmological media. In this background, we discussed the separation of the Klein-Gordon equation and then solved its radial part with numerical techniques to obtain QNMs and with analytical methods in terms of the general Heun functions to obtain the Hawking-Unruh and QBS spectra, with the angular part being solved in terms of the spherical harmonic functions under the assumption of slow rotation.

Specifically, we have exploited an adaptation of the WKB approximation to calculate the scalar QNMs of LTABHs. The mode behavior is quite similar to that of slowly-rotating Kerr BHs, with the Zeeman-like splitting affecting corotating and counterrotating QNMs with opposite signs of $m$. We have further analytically obtained the Hawking-Unruh radiation. Although superradiance is not achieved at the slow rotation limit, it contributes by decreasing the frequency content of the exponential argument in Eq.~(\ref{eq:rad_spec_LTABH}). Thus, the spectrum of Hawking-Unruh radiation will be higher (in modulus) than the ones concerning the nonrotating case. Finally, we obtained analytically the spectrum of QBSs of massless scalars in the LTABHs, as well as their respective radial and angular eigenfunctions which are well behaved at the boundaries. The slow rotation, encoded on the angular momentum $a/M$, gives rise to a nonzero oscillatory frequency, therefore the decay is no longer overdamped (for $\xi > 5$) as in the nonrotating case. The sign of the real part depends on the sign of the magnetic quantum number $m$. On the other hand, the decay rate of QBSs is mildly affected by the increment of rotation. Thus, from an experimental point of view, we expect to find, in principle, particles describing spirals as they go into the fluid flow sink. 

The overall analysis of QNMs and QBSs leads to the important conclusion that the LTABH is modally stable under scalar perturbations, since the imaginary parts of all of the frequencies are always negative, which translates to monotonous decay of perturbations in our harmonic expansion convention. Due to the similarities of slowly-rotating Kerr and LTABH QNMs it is tantalizing to extrapolate that by increasing the rotation to $a/M \rightarrow 1$ the spacetime may remain stable, though the effect of superradiance may change the picture.

As a future perspective, it is possible to extend our results in order to obtain a new acoustic curved BH embedded in Kerr spacetime, without the slow rotation assumption. Regarding observational prospects, condensed matter systems, and in particular Bose-Einstein condensates, provide one of the most promising experimental scenarios in which one can comprehend the physical aspects of superfluids at low temperatures. We believe that these systems may also provide a reasonable framework to observe the Lense-Thirring precession in an acoustic effective geometry described by a slowly-rotating fluid \cite{Chakraborty:2015ioa,Solnyshkov:2018dgq}. Astrophysical BHs with accretion disks are considered as the only classical analog system found in nature. Accreting BHs play a central role in explaining active galactic nuclei and possess both event and acoustic horizons. These systems can be used to study the total Hawking temperature with contributions from both the relativistic and the sonic sector. Recent studies have shown that the analog acoustic Hawking temperature dominates over the actual one in accreting BHs \cite{Das:2004wf,Das:2004zm,Abraham:2005ah}, therefore the transonic BH accretion process provides a test bed for the study of the properties of both horizons, as well as their astrophysical and observational competition. Our model spacetime may be used to interpret the effect of transonic accretion onto slowly-rotating BHs and place bounds on the behavior of the accreting fluid and Hawking radiation through the tuning parameter $\xi$. Besides matter accretion disks, BHs may be surrounded by dark matter halos \cite{Cardoso:2021wlq} consisting of quantum superfluids \cite{Berezhiani:2015bqa}. The transonic accretion and condensation of dark matter superfluids also provide a natural realization of analog gravity with $\xi$ providing a potential tuning parameter for the radial velocity of dark matter accretion.

\begin{acknowledgments}
H.S.V. is funded by the Alexander von Humboldt-Stiftung/Foundation (Grant No. 1209836). This study was financed in part by the Coordena\c c\~{a}o de Aperfei\c coamento de Pessoal de N\'{i}vel Superior - Brasil (CAPES) - Finance Code 001.
\end{acknowledgments}

\appendix

\section{Fluid flow with circulation}\label{AppendixA}
In this Appendix, we obtain an effective metric describing a slowly-rotating acoustic BH spacetime by directly extending the procedure of Ge \textit{et al.} \cite{Ge:2019our}. First, we present the highlights of their work. The background spacetime metric is fixed as a static one, namely
\begin{equation}
ds^{2}=g_{tt}\ dt^{2}+g_{rr}\ dr^{2}+g_{\vartheta\vartheta}\ d\vartheta^{2}+g_{\phi\phi}\ d\phi^{2}.
\label{eq2X:background_SBH_metric}
\end{equation}
In the Madelung representation (see Eq.~(\ref{eq:Madelung_representation})), the effective metric $\mathcal{G}_{\mu\nu}$ can be written as
\begin{equation}
	\mathcal{G}_{\mu\nu}=\frac{c_s}{\sqrt{c^2_s-v_{\mu}v^{\mu}}}
\begin{pmatrix}
g_{tt}(c^2_s- v_i v^i) & \vdots & -v_{t}v_{i}\cr
\cdots\cdots\cdots\cdots & \cdot &\cdots\cdots\cdots\cdots\cdots\cdots\cr
 -v_{i}v_{t} & \vdots & g_{ii}(c^2_s-v_\mu v^\mu)\delta^{ij}+v_i v_j\cr
\end{pmatrix}\!.
\label{eq2X:effective_metric_SABH}
\end{equation}
For the components of the fluid four-velocity, we can assume the following: $v_{t} \neq 0$, $v_{r} \neq 0$, $v_{\vartheta} = 0$, and $v_{\phi} \neq 0$ such that $v_{\phi}^{2} \approx 0$. We can also consider that $g_{rr}g_{tt}=-1$. Then, the new coordinate transformations are given by
\begin{eqnarray}
dt & \rightarrow & dt+\frac{v_{r}v_{t}}{g_{tt}(c_{s}^{2}-v_{i}v^{i})}dr,\\
d\phi & \rightarrow & d\phi-\frac{v_{r}v_{\phi}}{g_{\phi\phi}(c_{s}^{2}-v_{i}v^{i})}dr.
\label{eq2X:coordinate_transformations}
\end{eqnarray}
Thus, we obtain the following line element describing a spherically symmetric (and slowly-rotating) acoustic BH,
\begin{equation}
ds^{2}=c_{s}\sqrt{c_{s}^{2}-v_{\mu}v^{\mu}}\biggl(\frac{c_{s}^{2}-v_{r}v^{r}}{c_{s}^{2}-v_{\mu}v^{\mu}}g_{tt}\ dt^{2}+\frac{c_{s}^{2}}{c_{s}^{2}-v_{r}v^{r}}g_{rr}\ dr^{2}+g_{\vartheta\vartheta}\ d\vartheta^{2}+g_{\phi\phi}\ d\phi^{2}-2\frac{v_{\phi}v_{t}}{c_{s}^{2}-v_{\mu}v^{\mu}}\ dt\ d\phi\biggr).
\label{eq2X:acoustic_metric}
\end{equation}
The radial and temporal components of the fluid four-velocity can be chosen as in Eqs.~(\ref{eq2:radial_fluid_component}) and (\ref{eq2:temporal_fluid_component}). On the other hand, for the angular component $v_{\phi}$, which introduces the rotation, we set
\begin{equation}
v_{\phi} \sim \frac{C}{r},
\label{eq2X:angular_fluid_component}
\end{equation}
where $C$ is a real constant related to the circulation of the fluid flow, which can be chosen as $C=2Ma\sin^{2}\vartheta$. Then, we get
\begin{equation}
v_{\phi}v_{t} \approx \frac{2Ma\sin^{2}\vartheta}{r}+\mathcal{O}\biggl(\frac{1}{r^{2}}\biggr).
\label{eq2X:angular_temporal_product}
\end{equation}
Therefore, the line element (\ref{eq2X:acoustic_metric}) can be written as
\begin{equation}
ds^{2}=\sqrt{3}c_{s}^{2}\biggl[-\mathcal{F}(r)\ dt^{2}+\frac{1}{\mathcal{F}(r)}\ dr^{2}+r^{2}\ d\vartheta^{2}+r^{2}\sin^{2}\vartheta\ d\phi^{2}-\frac{4Ma\sin^{2}\vartheta}{r}\ dt\ d\phi\biggr],
\label{eq2X:LTABH_metric}
\end{equation}
which is the same effective metric given by Eq.~(\ref{eq2:LTABH_metric}), though in this case the slow rotation is due to the angular component of the four-velocity given in terms of a real constant related to the circulation of the fluid flow, which can be understood as a parameter in the lab framework.

%
%
\end{document}